\documentclass[preprint]{aastex}

\input psfig

\def\kms{{\,{\rm km}\,{\rm s}^{-1}}}
\def\kmsm{{\,{\rm km}\,{\rm s}^{-1}\,{\rm Mpc}^{-1}}}

\begin{document}

\title{The new two-image gravitational lens system CLASS B2319+051}

\author{D. Rusin, D.R. Marlow}
\affil{Department of Physics and Astronomy, University of Pennsylvania, 
209 S. 33rd St., Philadelphia, PA 19104-6396}

\author{M. Norbury, I.W.A. Browne, N. Jackson, P.N. Wilkinson} 
\affil{Jodrell Bank Observatory, University of Manchester, Macclesfield,
Cheshire SK11 9DL, UK}

\author{C.D. Fassnacht, S.T. Myers}
\affil{National Radio Astronomy Observatory, P.O. Box 0, Socorro, NM 87801}

\author{L.V.E. Koopmans\altaffilmark{1}, R.D. Blandford, T.J. Pearson,
A.C.S. Readhead} 
\affil{California Institute of Technology, 105-24, Pasadena, CA 91125}
\altaffiltext{1}{and Kapteyn Astronomical Institute, Postbus 800, NL-9700 AV
Groningen, Netherlands}

\author{A.G. de Bruyn\altaffilmark{1}}
\affil{Netherlands Foundation for Research in Astronomy, Postbus 2, NL-7990 AA
Dwingeloo, Netherlands}

\begin{abstract}

We report the discovery of a new two-image gravitational lens system from the
Cosmic Lens All-Sky Survey, CLASS B2319+051. Radio imaging with the Very Large
Array (VLA) and Multi-Element Radio-Linked Interferometer Network (MERLIN)
shows two compact components with a flux density ratio of $\simeq$ 5:1,
separated by $1\farcs36$. Observations with the Very Long Baseline Array
(VLBA) resolve each of the radio components into a pair of parity-reversed
subcomponents. Hubble Space Telescope (HST) observations with the
Near-Infrared Camera and Multi-Object Spectrometer (NICMOS) show a bright
elliptical galaxy (G1) coincident with the radio position, and a second
irregular galaxy (G2) $3\farcs4$ to the northwest. Previous spectroscopic
studies have indicated that these galaxies are at different redshifts ($z_{G1}
= 0.624$, $z_{G2}= 0.588$). Infrared counterparts to the lensed radio
components are not detected in the NICMOS image, and the source redshift has
not yet been determined. Preliminary mass modeling based on the VLBA
subcomponent data indicates that the lensing potential includes a strong
external shear contribution. A VLA monitoring program is currently being
undertaken to measure the differential time delay.

\end{abstract}


\section{Introduction} \label{sec:intro}

The Cosmic Lens All-Sky Survey (CLASS; Myers et al.\ 1995, 1999) seeks to
discover new cases of gravitational lensing among flat-spectrum radio sources,
and ultimately produce the largest and best studied sample of radio-selected
lenses. CLASS builds upon the success of the Jodrell-VLA Astrometric Survey
(JVAS; Patnaik et al.\ 1992; Browne et al.\ 1998; Wilkinson et al.\ 1998; King
et al.\ 1999) and extends the search to weaker flux densities. The
gravitational lens systems discovered by JVAS and CLASS are powerful tools for
investigating a wide range of astrophysical and cosmological problems. First,
arcsecond-scale lenses directly probe the inner several kiloparsecs of
galaxies at intermediate redshift (Kochanek 1991), and can place vital
constraints on their mass distributions. These studies indicate that the mass
profiles of early-type lensing galaxies are close to isothermal (e.g.,
Kochanek et al.\ 1995; Cohn et al.\ 2001; Rusin \& Ma 2001).  Second, measured
time delays between the images of a lensed source, when combined with a
well-constrained mass model, allow for a determination of the Hubble constant
(Refsdal 1964). Thus far time delays have been measured for seven
gravitational lens systems (Schechter et al.\ 1997; Kundic et al.\ 1997;
Lovell et al.\ 1998; Wisotzki et al.\ 1998; Biggs et al.\ 1999; Fassnacht et
al.\ 1999b; Koopmans et al.\ 2000) and favor a Hubble constant of $H_0 = 74
\pm 8 \kmsm$, assuming a flat cosmological model with $\Omega_{\Lambda} = 0.7$
(Koopmans \& Fassnacht 1999).  Third, the lensing rate in a systematic survey
can place upper limits on the cosmological constant (Turner, Ostriker \& Gott
1984; Turner 1990). Recent analyses favor $\Omega_{\Lambda} \leq 0.65$ for
flat cosmologies (Kochanek 1996; Falco, Kochanek \& Mu\~noz 1998; Quast \&
Helbig 1999).

Sources in the first two phases of CLASS were selected from the 87GB 5 GHz
catalog (Gregory \& Condon 1991), with $S_{5}\geq 25$ mJy and spectral index
$\alpha\geq-0.5$ (where $S_{\nu} \propto \nu^{\alpha}$) between 5 GHz and the
327 MHz Westerbork Northern Sky Survey (WENSS; Rengelink et al.\ 1997) or the
365 MHz Texas Survey (Douglas et al.\ 1996). Recently, the sources in the
CLASS sample were reselected using the 5 GHz GB6 catalog (Gregory et al.\
1996) and the 1.4 GHz NRAO VLA Sky Survey (NVSS; Condon et al.\ 1998), with a
5 GHz cutoff of 30 mJy and spectral index $\alpha\geq-0.5$ between 1.4 and 5
GHz. Spectral selection helps make CLASS a powerful and efficient lens
survey. The flat-spectrum CLASS sample is dominated by sources with
intrinsically compact morphologies, which eases the identification of lenses
and simplifies calculations of the statistical lensing rate. Furthermore,
compact sources tend to be variable, thereby making possible the measurement
of time delays from the lightcurves of lensed images.

The CLASS sample was observed using the Very Large Array (VLA) in A
configuration at 8.4 GHz, which offers a resolution of $\simeq 250$
mas. Observations of 3C286 were used to set the flux density scale. The VLA
data were calibrated in the Astronomical Image Processing System (AIPS), and
mapped using an automated script within the imaging package DIFMAP (Shepherd
1997). The data were then modeled with Gaussian components, which provides a
quantitative description of the observed radio morphology. Sources modeled
with multiple compact components are selected as preliminary lens
candidates. These candidates are followed up with high resolution radio
observations using the Multi-Element Radio-Linked Interferometer Network
(MERLIN; resolution of $\simeq 50$ mas at 5 GHz), and then the Very Long
Baseline Array (VLBA; resolution of $\simeq 5$ mas at 5 GHz) for the few
surviving sources. The vast majority of the lens candidates are rejected on
surface brightness and morphological grounds, and are instead shown to be
core-jet sources. Candidates survive the radio filter if their components have
compact or correlated structure at the milliarcsecond scale. These are then
followed up further with optical and/or near-infrared imaging and
spectroscopy.

CLASS survey observations of over 13000 sources were conducted in four phases
(CLASS 1--4) from spring 1994 to summer 1999, and are now complete. The first
phase of CLASS observations (CLASS--1) has yielded five new lens systems:
B0128+437 (Phillips et al.\ 2000), B0712+472 (Jackson et al.\ 1998), B1600+434
(Jackson et al.\ 1995), B1608+656 (Myers et al.\ 1995) and B1933+503 (Sykes et
al.\ 1998).  The second series of observations (CLASS--2) has produced another
five lenses: B0739+366 (Marlow et al.\ 2001), B1127+385 (Koopmans et al.\
1999), B1555+375 (Marlow et al.\ 1999b), B2045+265 (Fassnacht et al.\ 1999a),
and the one presented in this paper, B2319+051.  Two additional lens systems
have recently been discovered during the third phase of CLASS observations
(CLASS--3): B1152+199 and B1359+154 (Myers et al.\ 1999; Rusin et al.\
2000). Radio follow-up observations of the remaining CLASS--3 and CLASS--4
candidates are nearly complete.

Here we report the discovery of a new two-image gravitational lens from
CLASS--2: B2319+051. In section 2 we describe radio observations of the system
with the VLA, MERLIN and VLBA. Section 3 presents near-infrared imaging with
the Hubble Space Telescope. Preliminary mass modeling of B2319+051 is
described in Section 4.  Section 5 summarizes our results and discusses future
work.

\section{Radio Observations} \label{sec:obs}

B2319+051 was observed on 1995 August 13 during the second phase of the CLASS
survey observations. The 8.4 GHz discovery snapshot map has an rms noise of
$330$ $\mu$Jy/beam and is displayed in Fig.~1. The source consists of two
compact components with flux densities of $27.4 \pm 0.3$ mJy (A) and $5.0 \pm
0.3$ mJy (B) in a north-south orientation, separated by $1\farcs36$. Follow-up
VLA 1.4, 5, 8.4 and 15 GHz A configuration observations were performed on 1999
July 29 to investigate the spectral properties of the radio components. The
flux density scale was set by observations of the calibrator source
J2355+498. The VLA data sets were calibrated in AIPS using the standard
procedure and analyzed in DIFMAP. In each case the visibility data were fit to
a pair of compact Gaussian components using several iterations of
model-fitting and phase-only self-calibration (solution interval of 0.5
min). The component flux densities are given in Table 1, and the radio spectra
are plotted in Fig.~2. The radio spectra exhibit striking similarity, with
overall spectral indices between 1.4 and 15 GHz of $\alpha_{1.4}^{15} = -0.66
\pm 0.01$ (A) and $\alpha_{1.4}^{15} = -0.61 \pm 0.05$ (B), respectively. This
argues against the identification of A and B as either two independent quasars
or components of a core-jet structure. However, nearly identical radio spectra
would be expected for images of a gravitationally lensed source.

MERLIN 5 GHz observations of B2319+051 were performed on 1996 December 26 for
a total integration time of 1 hr, and again on 1998 February 24 for 8.5 hr to
improve the surface brightness sensitivity. Observations of 3C286 were used to
set the flux density scale. The data were calibrated in AIPS and imaged in
DIFMAP by repeating a cycle of cleaning and phase-only self-calibration,
starting with long solution intervals ($40$ min) and gradually decreasing to a
minimum interval of $2$ min. Once the model had sufficiently converged, an
amplitude self-calibration was performed using a solution interval of 30 min.
The final map has an rms noise of $70$ $\mu$Jy/beam and is shown in Fig~3. The
data were modeled by two compact Gaussian components with flux densities of
$44.0 \pm 0.1$ mJy (A) and $8.8 \pm 0.1$ mJy (B). No further emission was
detected down to the $3\sigma$ level of the maps. The compactness of each
radio component in the MERLIN map offers further evidence for the lensing
hypothesis.

VLBA 5 GHz observations of B2319+051 were performed on 1997 August 3.  The
observations were obtained using five separate snapshots, each of seven
minutes duration, over a range of hour angles to synthesize $uv$ coverage. The
system was reobserved at 5 GHz on 2000 September 18 for 11 hr to improve the
sensitivity. In each case, phase referencing was implemented using the nearby
calibrator source J2320+052. Fringe fitting was performed on J2320+052 and the
solutions were transferred directly to B2319+051. The calibrated data were
then imaged in DIFMAP using several iterations of cleaning and phase-only
self-calibration, with a solution interval of $5$ min. Maps from the deep
observation are presented in Fig.~4 and have an rms noise of $50$
$\mu$Jy/beam. Each of the radio components is resolved into a pair of compact
subcomponents (A1 and A2, B1 and B2), indicative of a ``core-knot'' source
morphology. In addition, we identify a very weak emission feature (A3) as the
beginning of a radio jet connecting A1 and A2. A corresponding feature is not
obvious in B, but its detection is greatly hindered by the decreased flux
density and angular size of the component.  B1 does however appear to be
slightly extended along the east-west direction, possibly signaling the
presence of this emission feature. While the maps we present employ natural
visibility weighting, the higher resolution obtained by switching to uniform
weighting is more than offset by the decreased sensitivity, and did not
improve the analysis. The visibility data were well fit by a total of five
compact ($< 1$ mas) Gaussian components, using iterated cycles of
model-fitting and phase-only self-calibration. The positions and flux
densities of the subcomponents are listed in Table 2.

Deep 1.7 GHz VLBA observations of B2319+051 were obtained on 1999 August
12. The total integration time was 12 hr. Phase referencing and fringe fitting
were performed with the nearby JVAS source J2322+082. The calibration and
imaging procedures were identical to those described above. The resulting maps
have an rms noise of 55 $\mu$Jy/beam and are displayed in Fig.~5. The
substructure observed in the 1.7 GHz maps matches that detected at 5 GHz. The
data were modeled by five Gaussian components that closely correspond to those
derived from the 5 GHz observations. Table 2 lists the flux densities of these
components. The residual map was searched for evidence of faint compact
emission features that could be associated with a third ``odd'' image of the
source (Rusin \& Ma 2001), but none were found down to the detection
threshold.

The milliarcsecond-scale substructure of the B2319+051 radio components
contains the unmistakable signature of gravitational lensing. First,
components A and B share nearly identical morphologies, as would be expected
for images of a common background source. Second, the relative inversion of
the radio components is a textbook example of lensing-induced parity reversal
(e.g., Schneider, Ehlers \& Falco 1992). Third, the brighter component (A)
covers a larger angular size, consistent with the conservation of surface
brightness by gravitational lensing. Taken together, these observations offer
compelling evidence that B2319+051 is a gravitational lens system.

Finally, we searched for changes in the substructure of the B2319+051 radio
images by comparing the relative positions of the core and knot subcomponents
in the 1997 and 2000 VLBA 5 GHz data. The likelihood of detecting evolving
superluminal jets is expected to be enhanced for lensed radio sources, as
magnification would increase the apparent transverse velocity. Jet evolution
may offer useful constraints on the mass model, and in particular on the local
magnification matrices, but this effect has not yet been observed in any
gravitational lens system. We find that the A1--A2 and B1--B2 separations in
the 1997 and 2000 data sets are each consistent to within $0.1$ mas.  Our
tests indicate that this is close to the reliability limit of DIFMAP
model-fitting.  Consequently, there is no evidence for evolving radio
substructure in B2319+051 at this time.

\section{HST Imaging and Astrometry}

Hubble Space Telescope (HST) observations of B2319+051 were obtained on 1998
May 30 with the Near-Infrared Camera and Multi-Object Spectrometer
(NICMOS). The F160W filter was used, which is centered at $1.6\mu$m and
corresponds roughly to ground-based $H$ band. The observations made use of the
NIC2 camera, with a detector scale of 75 mas/pixel and a field-of-view of
$19\farcs2\times19\farcs2$. The total exposure time was 2624 sec. The data
were subjected to the standard NICMOS calibration pipeline, involving bias and
dark current subtraction, flat-field correction, cosmic ray removal and
photometric calibration.  The final image is displayed in Fig.~6a. A bright
elliptical galaxy (G1) is observed at the position of the radio system, but no
counterparts to the radio images are detected. This suggests a very
optically-faint lensed source, similar to B1127+385 (Koopmans et al.\ 1999) or
B1933+503 (Marlow et al.\ 1999a). The observation of a galaxy close to the
expected position does, however, satisfy an important criterion of the lensing
hypothesis. A second, irregular galaxy (G2) lies $3\farcs4$ to the northwest
of G1.  A contour plot of the NICMOS image (Fig.~6b) clearly shows that G2 is
composed of two subcomponents (G2a and G2b).

Photometry and relative astrometry were performed on G1 and G2 (Table 3).  The
elliptical surface brightness profile of G1 has an axis ratio of $0.63\pm0.02$
(at the outmost isophote) and a position angle of $52^{\circ} \pm
5^{\circ}$. The integrated F160W magnitudes are 18.2 and 19.1 for G1 and G2,
respectively, with an uncertainty of $\pm 0.15$. Recent spectroscopic
observations using the Low Resolution Imaging Spectrograph (LRIS) on the
W. M. Keck telescope have demonstrated that the galaxies are not physically
associated: $z_{G1} = 0.624$ and $z_{G2} = 0.588$ (Lubin et al.\ 2000). The
spectrum of G1 is consistent with an early-type galaxy. G1 has approximately
the luminosity of an E/S0 $L*$ galaxy, including a K-correction but no
evolutionary correction (Poggianti 1997). The spectrum of G2 shows strong
Balmer absorption lines which, when combined with the irregular morphology,
suggest an interacting or merging system.

The absence of detectable counterparts to the lensed radio components means
that we cannot determine the positions of the galaxies relative to the images
using any one data set. We therefore attempted to extract the absolute
coordinates of G1 from the Keck observations of Lubin et al.\
(2000). Reference stars were selected from the USNO A2.0 Catalog (Monet et
al.\ 1996) and the astrometric solutions were calculated using Judy Cohen's
{\tt coordinates} program, which takes into account the distortions introduced
by the LRIS optics.  Eight unsaturated reference stars were selected for the
program, and the estimated final rms position error is
$0\farcs45$. Unfortunately, the saturation of many reference stars in the Keck
image makes sub-pixel accuracy impossible with the current data set. With the
solutions from the {\tt coordinates} program, we have determined the position
of G1 to be RA 23 21 40.817 Dec +05 27 36.57 (J2000). Based on the VLBA
positions of the lensed radio components, this would place G1 at ($0\farcs27$,
$-0\farcs66$) relative to A1, roughly between the two lensed images.

\section{Mass Modeling} \label{sec:sys}

B2319+051 presents a greater modeling challenge than many two-image
gravitational lens systems due to the lack of detected optical/infrared
counterparts to the lensed radio components. Though the astrometry is not
adequate to robustly fix the position of the lensing galaxy at this time, the
milliarcsecond-scale radio substructure offers a sufficient number of
constraints to test simple mass models. In this analysis we use the eight
coordinates ($x$, $y$) of A1, A2, B1 and B2, as well as the two 5 GHz flux
density ratios $r_{1}=|S_{B1}/S_{A1}|$ and $r_{2}=| S_{B2}/S_{A2}|$. Modeling
was performed with an image plane minimization (Kochanek 1991), which
optimized the fit statistic
\begin{equation}
\chi^{2} = \sum_{i=A1,A2,B1,B2} \left[ \frac { (x_{i}' - x_{i})^{2}}{\Delta
x_{i}^{2}}  +  \frac { (y_{i}' - y_{i})^{2}}{\Delta
y_{i}^{2}}     \right] +
\sum_{i=1,2}\frac{(r_i' -r_i )^{2}}{\Delta r_i^{2}}
\end{equation}
where primed quantities are model-predicted and unprimed quantities are
observed. We assumed a tolerance of $\Delta x = \Delta y = 0.1$ mas on the
image positions and $5\%$ on each of the flux densities ($\Delta r \simeq
7\%$) to account for possible modeling errors and source variability. A flat
$\Omega_{\Lambda}=0.7$ universe with $H_{0} = 100h \kmsm$ and a source
redshift of $z=1.5$ were assumed for all calculations.

We first modeled B2319+051 using a singular isothermal ellipsoid mass
distribution (SIE; Kormann, Schneider \& Bartelmann 1994) with surface density
\begin{eqnarray}
\Sigma(x,y) = {\sigma^2 \over 2 G} {\sqrt{f} \over \sqrt{x^2 + f^2 y^2}}
\end{eqnarray}
where $\sigma$ is the line-of-sight velocity dispersion and $f$ is the
projected axial ratio. Including the four parameters required to describe the
unlensed source subcomponents, the model has nine free
parameters. Consequently, the number of degrees of freedom (NDF) is one. Not
only does the model offer a rather poor fit to the radio data ($\chi^2$/NDF $
= 3.1$), but the mass distribution is predicted to lie at a position angle of
$\simeq -30^{\circ}$, nearly orthogonal to that of the observed surface
brightness of G1. This is unlikely to be a consistent scenario, as modeling
studies have demonstrated that lensing mass distributions are typically well
aligned with the light (Keeton, Kochanek \& Falco 1998). Fixing the mass
distribution at the observed galaxy position angle of $52^{\circ}$ leads to an
unacceptable fit. Even when the coordinate uncertainties are relaxed to a very
liberal $1$ mas, $\chi^2$/NDF $= 14$ for NDF $= 2$, with mismatch of the
subcomponent positions dominating the fit statistic. We interpret the poor
performance of the single galaxy model as evidence that external shear is
significantly influencing the lensing potential. Because the SIE model
requires an ellipticity oriented at $\simeq -30^{\circ}$, galaxies along this
axis are likely to be responsible for the shear. Lubin et al.\ (2000) detect
several galaxies, including G2, to the northwest of the system. These are
likely candidates for the perturbing mass.

We therefore expanded the lens model to include an external shear field of
constant direction and magnitude. To reduce the number of free parameters and
ensure a constrained model, we set the position angle and axial ratio of the
SIE to the observed values of the G1 surface brightness. The SIE + shear model
provides a much improved fit to the radio data ($\chi^2$/NDF $ = 0.45$ with
NDF = 1). The optimized model parameters are listed in Table 4. The preferred
shear field has a magnitude of $\gamma = 0.14$ and a position angle of
$-22^{\circ}$. G2 is unlikely to account for all of the model-predicted shear,
unless it has a much higher mass-to-light ratio than G1. (An isothermal mass
distribution at $3\farcs4$ requires a velocity dispersion of $\simeq 260 \kms$
to produce a shear field with $\gamma = 0.14$.) The position of the lensing
galaxy ($0\farcs21$, $-0\farcs79$) agrees well with that derived from our LRIS
astrometry. The predicted time delay between components A and B is $\simeq
18h^{-1}$ days for a flat $\Omega_{\Lambda} = 0.7$ cosmology.

\section{Summary and Future Work} \label{sec:fut}

VLA and MERLIN observations of CLASS B2319+051 show two compact radio
components separated by $1\farcs36$. Both components have virtually identical
radio spectra between 1.4 and 15 GHz, consistent with them being images of a
single background source. VLBA 5 GHz observations resolve each of the
components into a pair of compact subcomponents. VLBA 1.7 GHz observations
confirm this finding. The similar morphologies, consistent angular sizes and
relative inversion of the radio components provide powerful evidence in
support of the lensing hypothesis.  HST/NICMOS imaging has revealed a bright
elliptical galaxy at the expected position, along with a second irregular
galaxy $3\farcs4$ to the northwest. Infrared counterparts to the radio
components were not detected in this observation. However, the radio spectra
and substructure along with the identification of a lensing galaxy argue that
B2319+051 is, indeed, a genuine gravitational lens system.

Preliminary mass modeling has demonstrated that an isolated galaxy is unable
to account for the VLBA substructure while remaining consistent with the
properties of the lensing galaxy G1. Because an isolated SIE is required to be
oriented nearly perpendicular to the observed surface brightness of G1, nearby
galaxies may be contributing significant ellipticity to the lensing
potential. The introduction of external shear allows for an excellent fit to
the data, even when the position angle and axial ratio of the SIE are fixed to
match the G1 surface brightness. The favored shear direction is consistent
with mass distributions located to the northwest of G1, as detected by Lubin
et al.\ (2000).  Considering our inability to robustly fix the position of the
primary lensing mass relative to the lensed images, it is quite remarkable
that the simple substructure in the B2319+051 radio components not only
strongly excludes an isolated galaxy model, but also selects a reasonable
direction for the external shear field.

Of paramount importance to future studies of B2319+051 is the detection of
optical/infrared counterparts to the lensed images. The positions of these
images relative to the lensing galaxy will provide an essential check for our
preliminary modeling attempt, while offering additional constraints for the
construction of more realistic mass models. The detection of these images
would also be a first step toward measuring the source redshift, which is a
vital ingredient for Hubble constant determination. B2319+051 is currently
being monitored as part of a large VLA program to measure time delays in
JVAS/CLASS gravitational lens systems.

\acknowledgements

We thank the staffs of the VLA, MERLIN and VLBA for their assistance during
our observing runs. The National Radio Astronomy Observatory is a facility of
the National Science Foundation operated under cooperative agreement by
Associated Universities, Inc. MERLIN is operated as a National Facility by the
University of Manchester, on behalf of the UK Particle Physics \& Astronomy
Research Council. This research used observations with the Hubble Space
Telescope, obtained at the Space Telescope Science Institute, which is
operated by Associated Universities for Research in Astronomy Inc. under NASA
contract NAS5-26555. D.R. acknowledges funding from the Zaccheus Daniel
Foundation. This work was supported in part by European Commission TMR
Programme, Research Network Contract ERBFMRXCT96-0034 ``CERES''.

\clearpage


\begin{table*}[ht!]
\begin{tabular}{ c c c c c}
\hline
\hline  
Comp & $S_{1.4}$ & $S_{5}$ & $S_{8.4}$ & $S_{15}$\\
\hline
A & $70.8 \pm 0.3$ &  $55.8 \pm 0.1$ & $25.5 \pm 0.1$ & $14.8 \pm 0.4$ \\
B & $14.5 \pm 0.3$ &  $10.8 \pm 0.1$ & $ 5.3 \pm 0.1$ & $ 3.4 \pm 0.4$ \\
\hline
\end{tabular}
\caption{VLA component flux densities (in mJy) at 1.4, 5, 8.4 and 15 GHz. Data
for the 1999 July 29 observation. Errors in the flux densities are taken to be
equal to the rms noise of the respective maps.}
\end{table*}

\bigskip
\bigskip

\begin{table*}[ht!]
\begin{tabular}{ccccc}
\hline\hline
Component &$\Delta\alpha$& $\Delta\delta$ &  $S_{5}$ (mJy) & $S_{1.7}$ (mJy)\\
\hline
A1 & $0$            & $0$             & 31.0  & 42.7\\
A2 & $-0\farcs0203$ & $-0\farcs0037$  &  8.3  & 14.4\\
A3 & $-0\farcs0030$ & $-0\farcs0024$  &  0.8  &  9.4\\
B1 & $+0\farcs0113$ & $-1\farcs3638$  &  6.2  &  9.2\\ 
B2 & $+0\farcs0188$ & $-1\farcs3638$  &  1.9  &  2.8\\ 
\hline
\end{tabular}
\caption{VLBA data for B2319+051. Positions are from the 18 September 2000 5
GHz observation, and are offset from RA 23 21 40.8015 Dec +05 27 37.2252
(J2000). Flux densities are listed for both the 5 GHz and 1.7 GHz VLBA
observations. Model-fitting errors are $\simeq \pm 0.1$ mas for the positions
of A1, A2, B1 and B2. The positional error for A3 is substantially larger
($\simeq 1$ mas). Uncertainties on the flux densities are $\simeq 5\%$.}
\end{table*}

\bigskip
\bigskip

\begin{table*}[ht!]
\begin{tabular}{ccccc}
\hline\hline
Component &$\Delta\alpha$& $\Delta\delta$ &
$m_{F160W}$ \\
\hline
G1 & 0 & 0 & 18.2\\
G2a& $-2\farcs257$ & $2\farcs421$   &\\
G2b& $-2\farcs957$&$1\farcs777$ &\raisebox{2.5ex}[0cm]{19.1}\\ 
\hline
\end{tabular}
\caption{NICMOS galaxy positions and integrated magnitudes for B2319+051. The
errors on the positions are $\pm0\farcs015$. The errors on the magnitudes are
$\pm0.15$.}
\end{table*}

\bigskip
\bigskip

\begin{table*}[ht!]
\begin{tabular}{cc}
\hline\hline
Parameter & Value \\
\hline
$f_{G1}$       & $0.63$\\
$\sigma_{G1}$      & $237.7$ km/s\\
$(x,y)_{G1}$       & ($+0\farcs2085$, $-0\farcs7891$)\\
$PA_{G1}$          & $+52^{\circ}$\\
$\gamma$      & 0.141\\
$PA_{\gamma}$      & $-22.4^{\circ}$\\
$(x,y)_{src1}$     & ($+0\farcs2809$, $-0\farcs7606$)\\
$(x,y)_{src2}$     & ($+0\farcs2787$, $-0\farcs7583$)\\
$\mu_{A1,A2}$ &  $+12.73$, $+11.82$\\
$\mu_{B1,B2}$ &  $-2.62$, $-2.61$\\
$\Delta t$ &  $17.8$ $h^{-1}\ $ days\\
$\chi^{2}$/NDF     & $0.45$\\
\hline
\end{tabular}
\caption{The best-fit SIE + shear model parameters for B2319+051. All values
assume a flat $\Omega_{\Lambda} = 0.7$ cosmology. Listed are the fixed surface
density axial ratio $f_{G1}$, velocity dispersion $\sigma_{G1}$, coordinates
$(x,y)_{G1}$ and fixed position angle $PA_{G1}$ for the SIE; magnitude
$\gamma$ and position angle $PA_{\gamma}$ for the external shear; recovered
source coordinates $(x,y)_{src}$, predicted magnifications $\mu$ and time
delay $\Delta t$. The image positions are reproduced almost exactly, and are
not listed.}
\end{table*}

\clearpage


\begin{figure*}
\figurenum{1}
\psfig{file=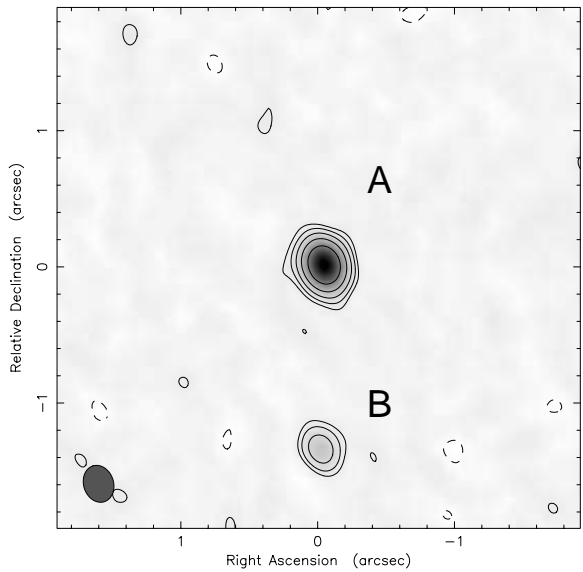,width=3in}
\caption{VLA 8.4 GHz discovery snapshot observation of B2319+051 taken
1995 August 29. The lowest contour is at $\pm 3\%$ of the map peak of $26.3$
mJy/beam, and contour levels increase by factors of 2. The synthesized beam is
$0\farcs276 \times 0\farcs219$ at $17.4^{\circ}$. The data have been naturally
weighted.}
\end{figure*}

\begin{figure*}
\figurenum{2}
\psfig{file=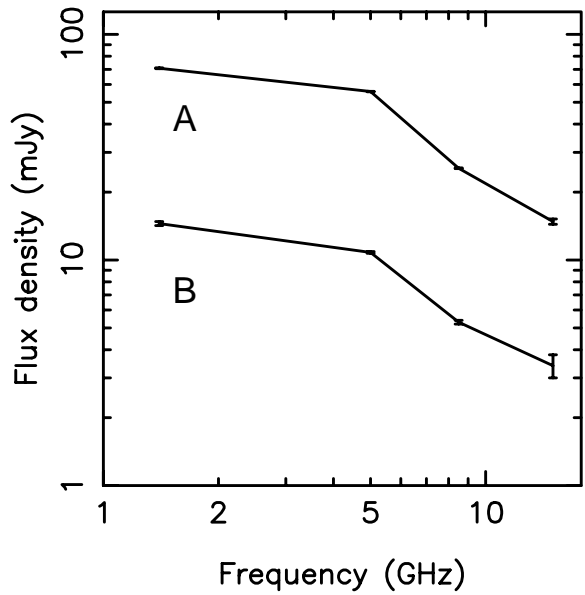,width=3in}
\caption{Component radio spectra based on the VLA 1.4, 5, 8.4 and 15 GHz data
of 1999 July 29.}
\end{figure*}

\clearpage

\begin{figure*}
\figurenum{3}
\psfig{file=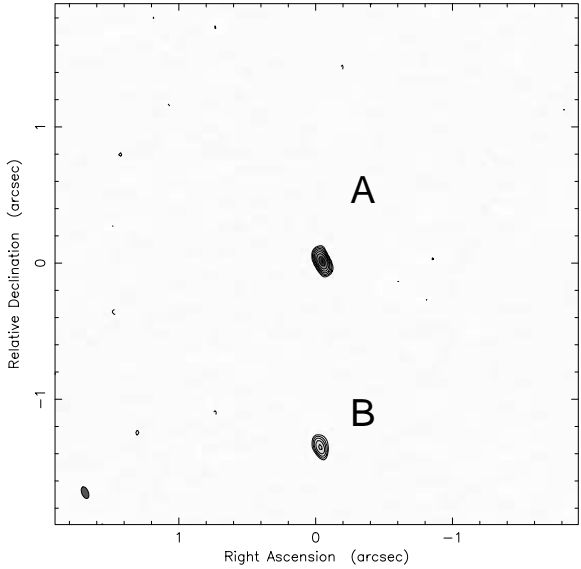,width=3in}
\caption{MERLIN 5 GHz observation of B2319+051 taken 1998 February 24.
The lowest contour is at $\pm 1\%$ of the map peak of $42.4$ mJy/beam, and
contour levels increase by factors of 2. The synthesized beam is $0\farcs091
\times 0\farcs048$ at $21.9^{\circ}$. The data have been naturally weighted.}
\end{figure*}

\begin{figure*}
\begin{tabular}{c c}
\psfig{file=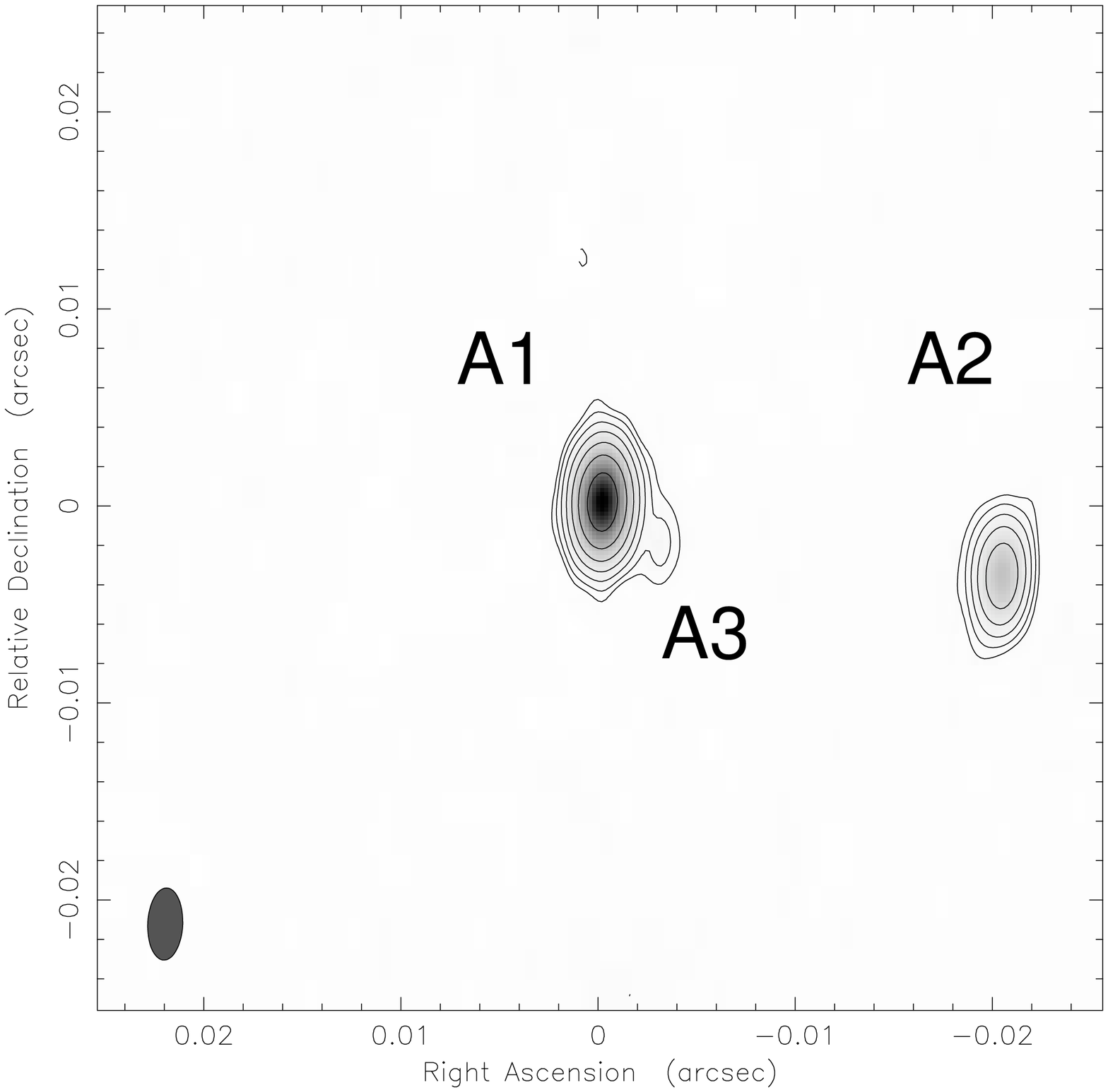,width=3in} & 
\psfig{file=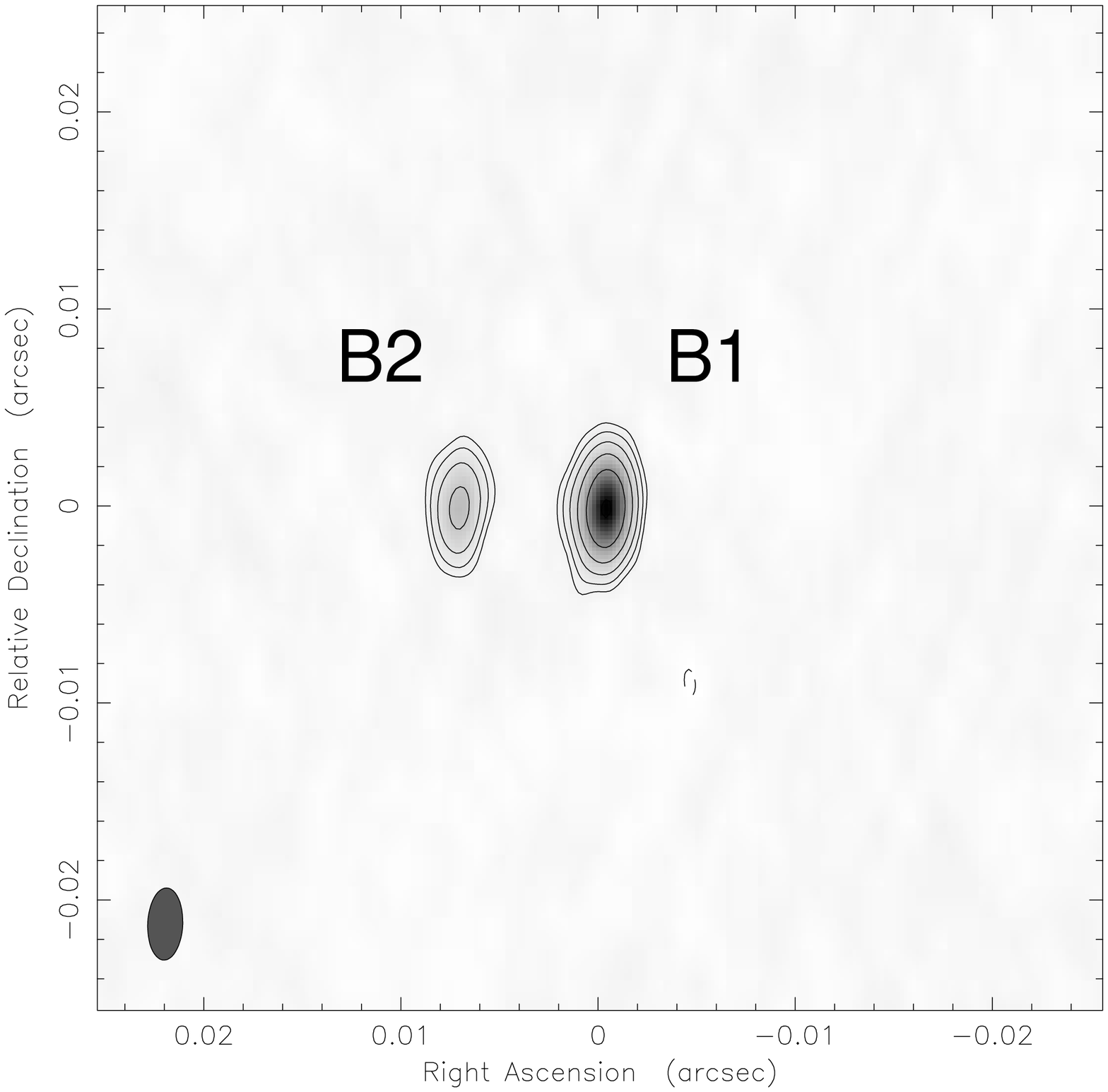,width=3in}\\
\end{tabular}
\figurenum{4}
\caption{VLBA 5 GHz observation of B2319+051 taken 2000 September 18. The beam
is $3.7 \times 1.8$ mas at $-2.5^{\circ}$. The data have been naturally
weighted. (a) Left: Component A. The lowest contour is at $\pm 1\%$ of the map
peak of $30.1$ mJy/beam, and contour levels increase by factors of 2.  (b)
Right: Component B.  The lowest contour is at $\pm 3\%$ of the map peak of
$5.9$ mJy/beam, and contour levels increase by factors of 2.}
\end{figure*}

\clearpage

\begin{figure*}
\begin{tabular}{c c}
\psfig{file=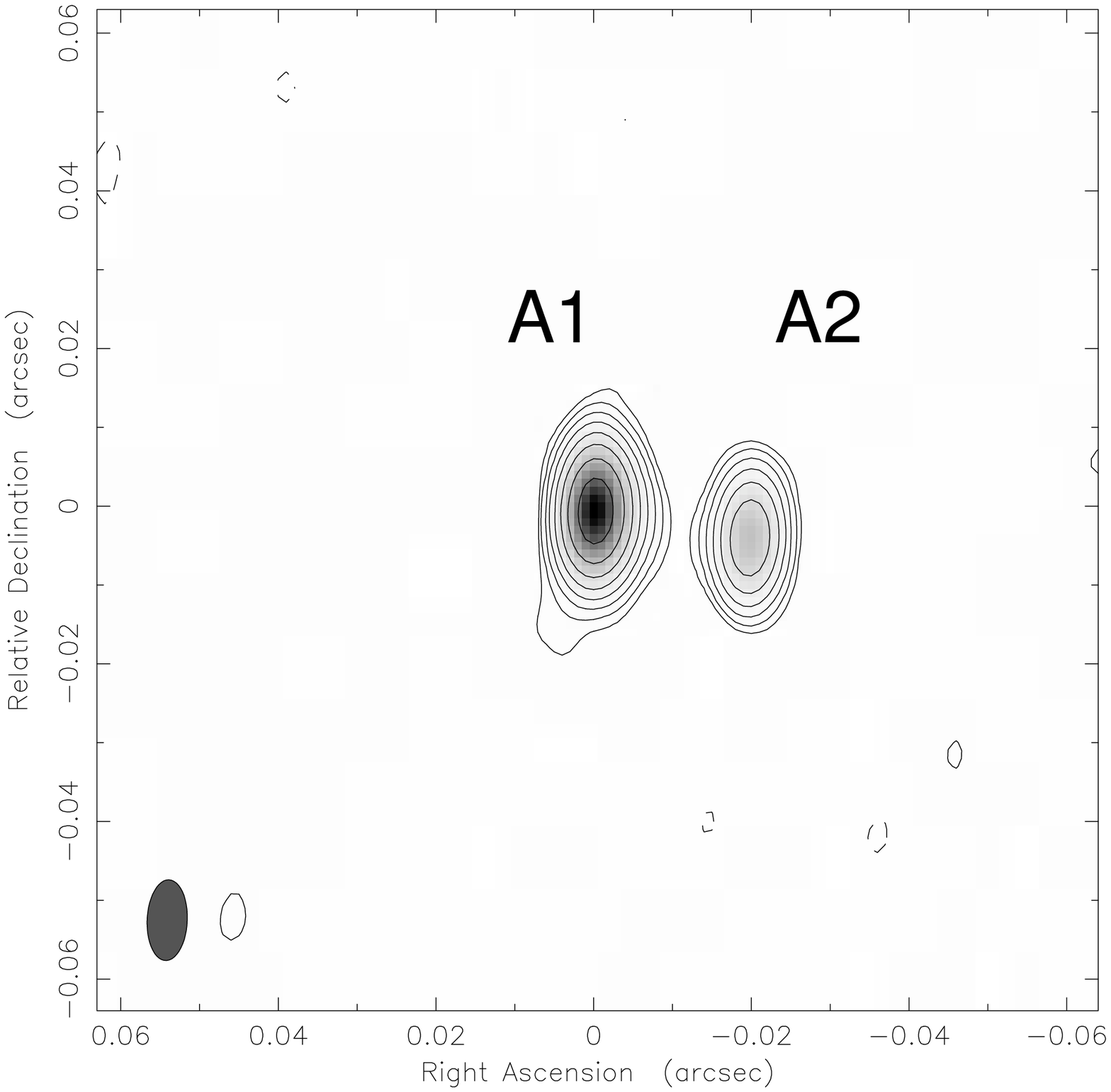,width=3in} & 
\psfig{file=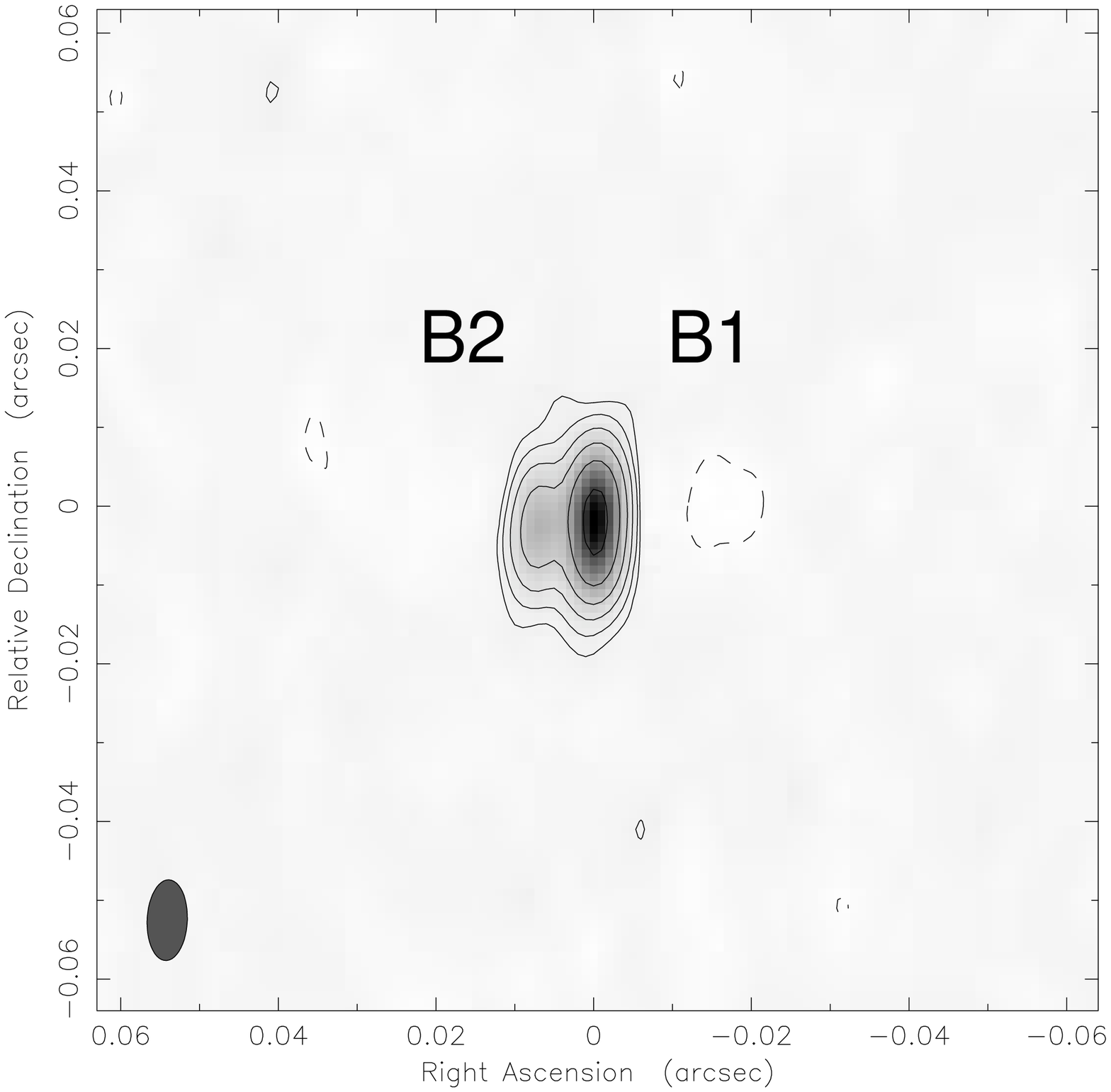,width=3in}\\
\end{tabular}
\figurenum{5}
\caption{VLBA 1.7 GHz observation of B2319+051 taken 1999 August 12. The beam
is $10.3 \times 5.1$ mas at $-2.6^{\circ}$. The data have been naturally
weighted. (a) Left: Component A. The lowest contour is at $\pm 0.5\%$ of the
map peak of $46.8$ mJy/beam, and contour levels increase by factors of 2. (b)
Right: Component B. The lowest contour is at $\pm 2.5\%$ of the map peak of
$6.3$ mJy/beam, and contour levels increase by factors of 2.}
\end{figure*}

\begin{figure*}
\begin{tabular}{c c}
\psfig{file=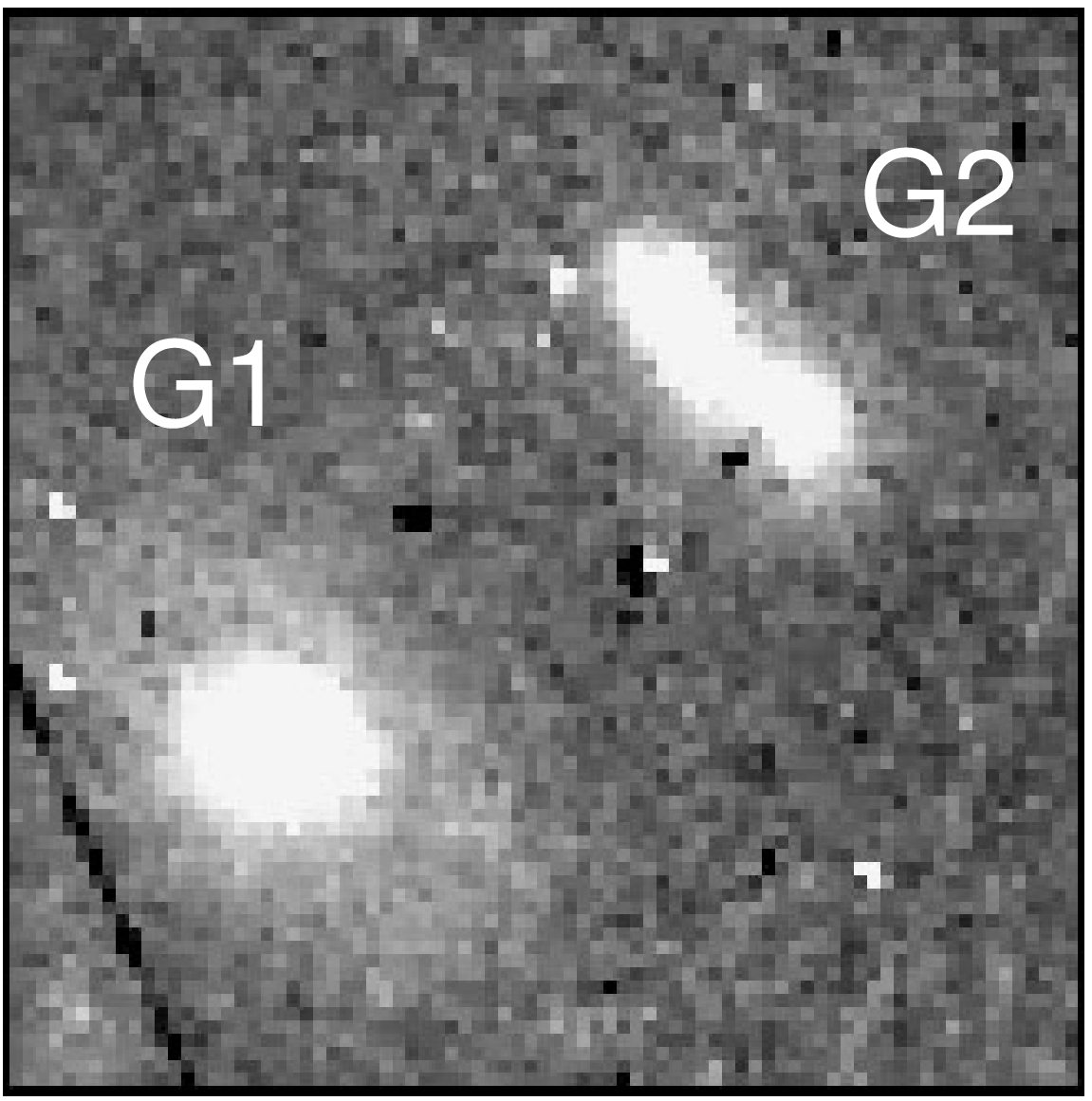,width=3in} & 
\psfig{file=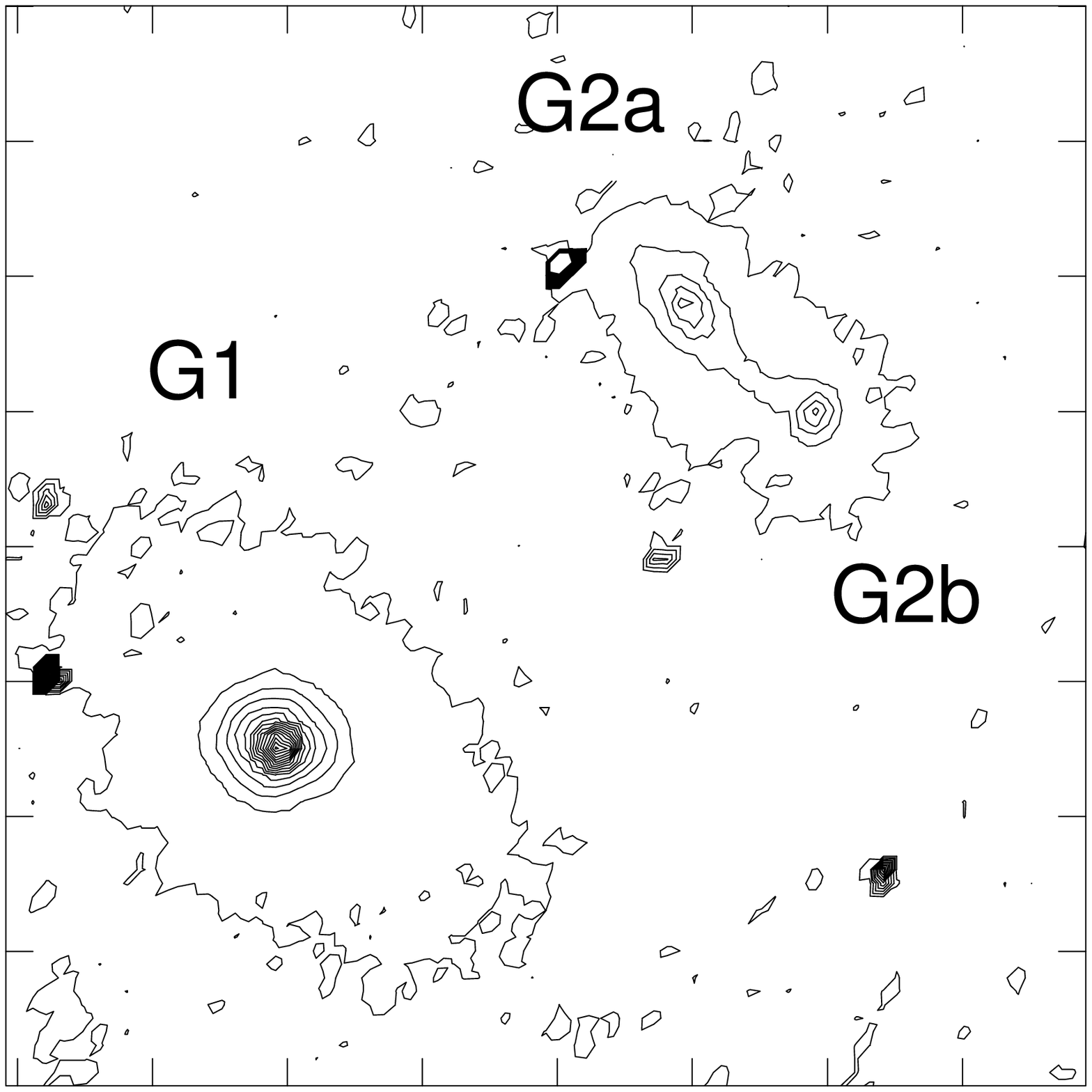,width=3in}\\
\end{tabular}
\figurenum{6}
\caption{NICMOS F160W images of B2319+051. North is up, east is left.  The
area shown is $6\farcs0 \times 6\farcs0$. (a) Left: Final map. (b). Right:
Contour map. The contour levels are (1, 2, 3, $\ldots$ 28) $\times$ rms noise
in the image. Note the absence of any counterparts to the lensed radio images,
and the two surface brightness peaks of G2. }
\end{figure*}

\clearpage

\begin{figure*}
\figurenum{7}
\psfig{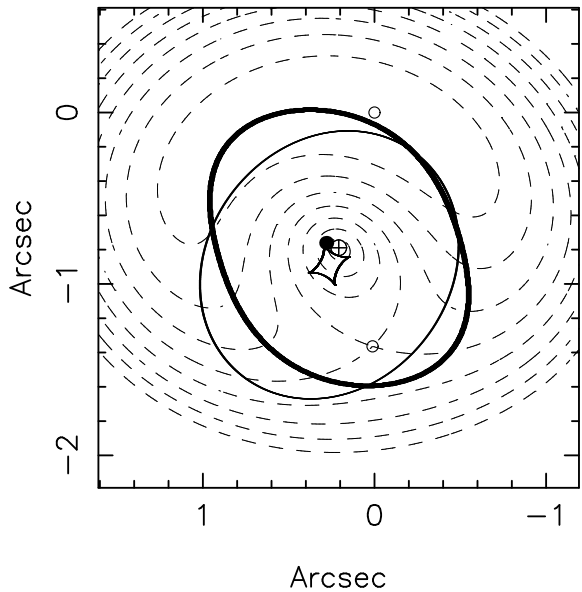} 
\caption{The critical curve (thick line) and caustics (thin lines) of the SIE
+ shear lens model. The filled circle marks the recovered source position.
The open circles indicate the positions of the images.  The center of the SIE
is marked by the cross-haired circle. Dashed lines denote contours of constant
time delay in increments of $4.45 h^{-1}$ days outward from the global minimum
at image A. The caustics are offset from the position of the lensing galaxy
due to our choice of ``centering'' the shear at (0,0). This has no effect on
the optimized model parameters or predicted time delay.}
\end{figure*} 


\clearpage

\begin{references}

\reference{0218delay}
Biggs, A.D., Browne, I.W.A., Helbig, P., Koopmans, L.V.E., Wilkinson, P.N., \&
Perley, R.A. 1999, MNRAS, 304, 349

\reference{browne98}
Browne, I.W.A., Wilkinson, P.N., Patnaik, A.R., \& Wrobel, J.M. 1998, MNRAS,
293, 257

\reference{cohn1933} 
Cohn, J.D., Kochanek, C.S., McLeod, B.A., \& Keeton, C.R.  2001, 
ApJ, in press (astro-ph/0008390)

\reference{nvss}
Condon, J.J., Cotton, W.D., Greisen, E.W., Yin, Q.F., Perley, R.A., Taylor,
G.B., \& Broderick, J.J. 1998, AJ, 115, 1693

\reference{douglas96} 
Douglas, J.N., Bash, F.N., Bozyan, F.A., Torrence, G.W., \& Wolfe, C. 1996,
AJ, 111, 1945

\reference{falco98}
Falco, E.E., Kochanek, C.S., \& Mu\~noz, J.A. 1998, ApJ, 494, 47

\reference{2045}
Fassnacht, C.D., et al.\ 1999a, AJ, 117, 658

\reference{1608monitor}
Fassnacht, C.D., Pearson, T.J., Readhead, A.C.S., Browne, I.W.A., Koopmans,
L.V.E., Myers, S.T., \& Wilkinson, P.N. 1999b, ApJ, 527, 498
 
\reference{87gb} 
Gregory, P.C., \& Condon, J.J. 1991, ApJS, 75, 1011
 
\reference{gb6}
Gregory, P.C., Scott, W.K., Douglas, K., \& Condon, J.J. 1996, ApJS, 103, 427

\reference{1600}
Jackson, N., et al. 1995, MNRAS, 274L, 25

\reference{0712}
Jackson, N., et al. 1998, MNRAS, 296, 483

\reference{KKF}
Keeton, C.R., Kochanek, C.S., \& Falco, E.E. 1998, ApJ, 509, 561

\reference{king99}
King, L.J., Browne, I.W.A., Marlow, D.R., Patnaik, A.R., \&
Wilkinson, P.N. 1999, MNRAS, 307, 225

\reference{koch91}
Kochanek, C.S. 1991, ApJ, 373, 354

\reference{koch95}
Kochanek, C.S. 1995, ApJ, 445, 559

\reference{koch96a}
Kochanek, C.S. 1996, ApJ, 466, 638

\reference{1127}
Koopmans, L.V.E., et al. 1999, MNRAS, 303, 727

\reference{1608delay}
Koopmans, L.V.E., \& Fassnacht, C.D. 1999, ApJ, 527, 513

\reference{1600delay}
Koopmans, L.V.E., de Bruyn, A.G., Xanthopoulos, E., \& Fassnacht, C.D. 2000,
A\&A, 356, 391

\reference{kormann}
Kormann, R., Schneider, P., \& Bartelmann, M. 1994, A\&A, 284, 285

\reference{0957delay}
Kundic, T., et al.\ 1997, ApJ, 482, 75

\reference{1830delay}
Lovell, J.E.J., Jauncey, D.L., Reynolds, J.E., Wieringa, M.H., King, E.A.,
Tzioumis, A.K., McCullough, P.M., \& Edwards, P.G. 1998, ApJ, 508L, 51

\reference{lubin.spec}
Lubin, L.M., Fassnacht, C.D., Readhead, A.C.S., Blandford, R.D., \& Kundic,
T. 2000, AJ, 119, 451

\reference{1933hst}
Marlow, D.R., Browne, I.W.A., Jackson, N., \& Wilkinson, P.N. 1999a, MNRAS,
305, 15

\reference{1555}
Marlow, D.R., et al. 1999b, AJ, 118, 654

\reference{0739}
Marlow, D.R., et al.\ 2001, AJ, 121, 619

\reference{monet}
Monet, D., et al.\ 1996, USNO-SA2.0 (U.S. Naval Observatory, Washington DC)

\reference{1608}
Myers, S.T., et al. 1995, ApJ, 447L, 5

\reference{myers99} 
Myers, S.T., et al. 1999, AJ, 117, 2565

\reference{patnaik92} 
Patnaik, A.R., Browne, I.W.A., Wilkinson, P.N., \& Wrobel, J.M. 1992, 
MNRAS, 254, 655

\reference{0128}
Phillips, P.M., et al.\ 2000, MNRAS, 319L, 7

\reference{pogg}
Poggianti, B.M. 1997, A\&AS, 122, 399

\reference{quast}
Quast, R., \& Helbig, P. 1999, A\&A, 344, 721

\reference{refsdal}
Refsdal, S. 1964, MNRAS, 128, 307

\reference{renge}
Rengelink, R.B., Tang, Y., de Bruyn, A.G., Miley, G.K., 
Bremer, M.N., Roettgering, H.J.A., \& Bremer, M.A.R. 1997, A\&AS, 124, 259

\reference{rusinetal2000}
Rusin, D., Hall, P.B., Nichol, R.C., Marlow, D.R., Richards, A.M.S., \& Myers,
S.T. 2000, ApJ 533L, 89
 
\reference{rusin.ma}
Rusin, D., \& Ma, C.-P. 2001, ApJ, 549L, 33

\reference{1115delay}
Schechter, P.L., et al. 1997, ApJ, 475L, 85

\reference{SEF}  
Schneider, P., Ehlers, J., \& Falco, E.E. 1992, {\em Gravitational Lenses}
(Berlin: Springer-Verlag)

\reference{difmap} 
Shepherd, M.C. 1997, in Astronomical Data Analysis Software 
and Systems IV, eds. G. Hunt \& H. E. Payne, (ASP Conference 
Series, v125) 77

\reference{1933}
Sykes, C.M., et al. 1998, MNRAS, 301, 310

\reference{TOG}
Turner, E.L., Ostriker, J.P., \& Gott, J.R. 1984, ApJ, 284, 1

\reference{turner}
Turner, E.L. 1990, ApJ, 365L, 43

\reference{wilkinson98}
Wilkinson, P.N., Browne, I.W.A., Patnaik, A.R., Wrobel, J.M., \& Sorathia, B.
1998, MNRAS, 300, 790

\reference{wisotzki.delay}
Wisotzki, L., Wucknitz, O., Lopez, S., \& Sorensen, A.N. 1998, A\&A, 339L, 73

 
\end{references}
\end{document}